\documentclass[runningheads]{llncs}
\PassOptionsToPackage{table}{xcolor}
\usepackage[table]{xcolor}
\definecolor{darkpastelgreen}{rgb}{0.01, 0.75, 0.24}
\usepackage[T1]{fontenc}
\usepackage{amsmath}
\usepackage{amsfonts}
\usepackage{cite}
\usepackage{booktabs}
\usepackage[pagebackref,breaklinks,colorlinks,citecolor=black, linkcolor=black, urlcolor=blue]{hyperref}
\usepackage{pgfplots}
\usepackage{pgfplotstable}
\usepackage{caption}
\usepackage{adjustbox}
\usepackage{pifont}
 \usepackage{multirow}
 \usepackage{cleveref}
\pgfplotsset{compat=1.18}

\usepackage[colorinlistoftodos]{todonotes}
\usepackage{graphicx}
\usepackage{color}

\begin{document}

\title{Unmasking Interstitial Lung Diseases:\\ Leveraging Masked Autoencoders for Diagnosis}

\titlerunning{Unmasking Interstitial Lung Diseases}
%


\author{Ethan Dack\inst{1} \and
Lorenzo Brigato\inst{1} \and
Vasilis Dedousis\inst{1} \and
Janine Gote-Schniering \inst{1} \and
Cheryl Magnin \inst{1} \and
Hanno Hoppe \inst{1, 2,3,4} \and
Aristomenis Exadaktylos \inst{2}
Manuela Funke-Chambour \inst{2} \and
Thomas Geiser \inst{2} \and
Andreas Christe \inst{2,3} \and
Lukas Ebner \inst{2,3} \and
Stavroula Mougiakakou \inst{1,2}}
\authorrunning{Dack, Brigato, et al.}

\institute{
University of Bern \and  
Bern University Hospital \and
DIPR \and
Campus Stiftung Lindenhof Bern (SLB)
 }

\maketitle              
\email{[ethan.dack, lorenzo.brigato, stavroula.mougiakakou]@unibe.ch}
\begin{abstract}

Masked autoencoders (MAEs) have emerged as a powerful approach for pre-training on unlabelled data, capable of learning robust and informative feature representations.
This is particularly advantageous in diffused lung disease research, where annotated imaging datasets are scarce. 
To leverage this, we train an MAE on a curated collection of over 5,000 chest computed tomography (CT) scans, combining in-house data with publicly available scans from related conditions that exhibit similar radiological patterns, such as COVID-19 and bacterial pneumonia. 
The pretrained MAE is then fine-tuned on a downstream classification task for diffused lung disease diagnosis. 
Our findings demonstrate that MAEs can effectively extract clinically meaningful features and improve diagnostic performance, even in the absence of large-scale labelled datasets. 
The code and the models are available here: \href{https://github.com/eedack01/lung_masked_autoencoder}{https://github.com/eedack01/lung\_masked\_autoencoder}.

\end{abstract}
\section{Introduction}

Artificial intelligence (AI) has demonstrated significant success in matching clinical experts' diagnostic performance for respiratory diseases, including infectious conditions, like COVID-19, and non-infectious ones like interstitial lung diseases (ILDs) \cite{01, 02}.
ILD diagnosis is difficult to classify as there are more than 200 different types and patterns. 
A crucial aspect in developing diagnostic tools focuses on imaging data, essential for identifying pathological patterns \cite{05}.
AI has been widely successful in other areas of diagnostic medicine, being implemented in daily practice across Europe \cite{04}, but this has been neglected in respiratory diseases such as ILD \cite{055}.
With traditional supervised approaches being the standard method of diagnosing, other methods, such as unsupervised and self-supervised approaches, are still widely unexplored in this area \cite{05}. 
Utilising autoencoders has been well established in assisting with diagnostic tasks \cite{12}.

Vision transformers (ViTs) have been extremely popular in medical-supervised learning tasks and have been applied to lung scans regarding ILDs and COVID-19 \cite{SAHA2024779, sabir2023fibrovit, covid-vit}. 
We aim to explore the use of unlabelled data to support the downstream task.
Self-supervised methods, such as contrastive learning, are challenging to train in 3D medical domains because it is shown they perform better on downstream tasks with larger batch sizes \cite{chen2020simpleframeworkcontrastivelearning, NEURIPS2022_db174d37, dack2023empirical}.
This demand significantly increases computational costs due to the large size of medical images, making it impractical for many researchers \cite{Huang2023}.
Still, input space reconstruction without masking fails to create uninformative latent features for downstream tasks \cite{balestriero2024learningreconstructionproducesuninformative}.
Naturally, the development of the ViT led to the masked autoencoder (MAE). 
These have shown great success in the non-medical domain \cite{he2021maskedautoencodersscalablevision}, and recent approaches have begun to introduce them in a medical setting \cite{zhou2023selfpretrainingmaskedautoencoders, chen2022maskedimagemodelingadvances, Kun_Training_MICCAI2024}. 
By making the task more difficult, masking input images compels the model to learn a deeper semantic understanding, resulting in more accurate reconstructions and more informative feature representations.
MAEs enable efficient, scalable training by extending ViT with unsupervised learning, offering a modern alternative to convolutional autoencoders while reducing overfitting. \cite{he2021maskedautoencodersscalablevision}
MAEs also allow us to pre-train on large amounts of unlabelled data and fine-tune for ILD tasks where data is traditionally not in abundance. 
We fine-tune our MAE for the difficult downstream task of ILD diagnosis.

Diagnosing interstitial lung diseases (ILDs) requires a multidisciplinary team and often a lung biopsy. 
However, radiologists can use CT scans to reduce the need for these invasive procedures by classifying patients into usual interstitial pneumonia (UIP), probable UIP, indeterminate UIP, or non-idiopathic pulmonary fibrosis (non-IPF) categories \cite{Lynch2018}.
AI-based CT analysis offers a reliable second opinion, particularly in settings where expert interpretation is limited, such as developing countries.
Due to the overlapping features observed in the intricate lung details between patients affected COVID-19 and those with ILDs, this research leverages the extensive imaging data collected during the pandemic to address the scarcity of data on ILD patients.
The nature of our tasks aligns well with utilising MAE models for learning representations for patients suffering from lung disease.
To summarise, the contributions of our work are as follows: 
\begin{itemize}
    \item[--] We leverage COVID-19 and ILD data to build a larger dataset for self-supervised training, enabling fine-tuning for respiratory diseases where data is traditionally limited. The model is released for research purposes.
    \item[--] We conduct an extensive analysis comparing established baselines with our pre-trained MAE to assess its effectiveness in classifying diffuse lung diseases.
\end{itemize}

\section{Method}

\subsection{Training strategy}
\subsubsection{MAE Pretraining} Following \cite{chen2022maskedimagemodelingadvances}, we use a ViT-Base (ViT-B) architecture, provided by the MONAI library \cite{monai}, as our encoder backbone, consisting of 12 layers, 12 attention heads, and a hidden dimension of 768. 
To handle 3D medical volumes, we adapt the input format accordingly.
All volumes are resized to a fixed size of 128$\times$128$\times$128, and we use a patch size of 16, yielding 512 non-overlapping patches. 
These patches are passed through the encoder to obtain patch embeddings, to which we add positional encodings. 
A subset of these embeddings is uniformly masked (75\%) and the masked positions are also augmented with positional embeddings.
The resulting sequence is passed through the decoder, and we compute the mean absolute error between the reconstructed patches and their original masked counterparts.

\begin{figure}[t]
\raggedright
\includegraphics[width=\linewidth, height=6cm, keepaspectratio]{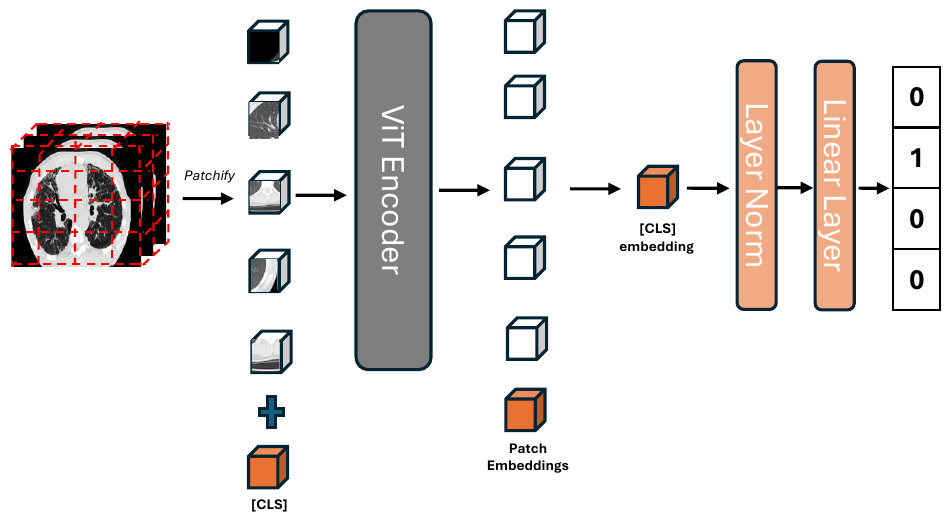}
\caption{\textbf{Finetuning strategy for ILD diagnosis.} The class token ($\texttt{[CLS]}$) is added to the patchified input and passed through the encoder. Only the class token is used after encoding, followed by normalization and a linear layer for classification using the supervised cross-entropy loss.}
\label{fig:finetuning-fig}
\end{figure}

\subsubsection{Fine-tuning for Classification} We replicate the experimental scenario from \cite{29,28}, solving two classification tasks, one multiclass and one binary. 
To fine-tune our MAE, we first patchify a volume, with the resulting tokens, we create an additional token known as the class token ($\texttt{[CLS]}$).
$\texttt{[CLS]}$ serves as a learnable summary representation of all other tokens, enabling efficient fine-tuning by providing a compact and consistent output.
As seen in Figure \ref{fig:finetuning-fig}, all tokens are passed through the encoder blocks of the MAE.
We select only the class token, pass this through a normalization and linear layer to solve the classification problem. 
For both tasks, we employ the standard cross-entropy loss function.
In the instance of binary classification, the classes are approximately in the ratio of 45:55, whereas the multiclass problem is more imbalanced.
To counter the imbalance problem, we calculate class weights such that $w_i= \frac{N}{n_i}$, where $w_i$ is the weight for class $i$, $N$ is the total number of samples, and $n_i$ is the number of samples in class $i$.
The weights are passed directly into the loss function.

\begin{table}[t]
\centering
\caption{\textbf{Number of CT scans per disease category.} Certified refers to a diagnosis assigned through board consensus. Uncertified lacks board consensus.}
\label{tab:disease-scans}
\setlength{\tabcolsep}{4pt}
\resizebox{\linewidth}{!}{
\begin{tabular}{l|ccccc}
\toprule
\textbf{Disease Category} & \textbf{Certified ILD} & \textbf{Uncertified ILD} & \textbf{COVID-19} & \textbf{Normal} & \textbf{Pneumonia} \\
\midrule
\textbf{Number of Scans} & 105 & 472 & 4,296 & 330 & 60 \\
\bottomrule
\end{tabular}
}
\end{table}

\begin{table}[h]
\centering
\caption{Breakdown of the Certified ILD cases.}
\label{tab:ild_dist}
\setlength{\tabcolsep}{4pt}
\resizebox{\linewidth}{!}{
\begin{tabular}{l|ccccc}
\toprule
\textbf{Disease Category} & \textbf{UIP} & \textbf{Probable UIP} & \textbf{Indeterminate for UIP} & \textbf{non-IPF}  \\
\midrule
\textbf{Number of Cases} & 27 & 21 & 18 & 39 \\
\bottomrule
\end{tabular}
}
\end{table}

\subsection{Datasets}

\subsubsection{Pre-training datasets}

To build a broad pre-training dataset for our MAE, we collected a large number of thoracic CT scans reflecting ILD-related imaging patterns. 
Given the overlap between COVID-19 and ILD features \cite{Durhan2023}, we included COVID-era scans to boost representation. 
Datasets focusing on lung nodules were excluded, as they do not reflect ILD abnormalities. 
A small number of normal scans were also included to assist the model in understanding the shape and patterns of the lung parenchyma.
For ILD-positive patients, we collected data from 360 individuals in our study and 176 from open-source platforms \cite{34, Depeursinge2012}. 
We further augmented the dataset with approximately 3,000 additional cases sourced from publicly available datasets \cite{Zaffino2021, Desai2020, MP-COVID-19-SegBenchmark, Morozov2020, An2020, stoic, Kiser2020, Oikonomou2021}, bringing the total to over 5,000 chest CT scans.
As shown in Table \ref{tab:disease-scans}, COVID-19 data constitutes the majority, followed by ILD cases. 
Our goal is to leverage the shared imaging characteristics of ILD and COVID-19 to enhance ILD diagnosis through a combined dataset.

\subsubsection{Fine-tuning datasets}

As seen in Table \ref{tab:disease-scans}, we have 105 CT scans which have been labelled according to the diagnostic criteria of the Fleischner Society\cite{Lynch2018}.
A distribution of the labels can be seen in Table \ref{tab:ild_dist}.
As stated, the dataset has some imbalance in the multiclass task, with the most numerous class having approximately double the number of cases of the least common one. 
Following previous works \cite{28,29}, the binary classification task is formed by combining the first two classes (“UIP” and “probable UIP”) into one category and the last two (“indeterminate for UIP” and “most consistent with non-IPF”) into another.

\subsubsection{Preprocessing} To ensure consistent preprocessing and avoid shortcuts during pre-training, we treat all scans uniformly.
First, all scans are converted to NIfTI format. 
We then resample each image to match the average voxel spacing across dimensions.
Utilising the open-source LungMask library \cite{lungmask}, we generate the lung mask, which acts as a bounding box for cropping the scan to zoom in on the lung parenchyma to avoid any irrelevant information on the outside of the image.
Finally, we apply min-max normalisation using Hounsfield unit bounds of -200 and 1200, scaling voxel values to the $[0, 1]$ range.

\subsection{Experimental Settings}

\subsubsection{Implementation and Hyperparameters} We summarise the experimental set-up in Table \ref{tab2}.
Where possible, we implement existing hyperparameters from previous papers \cite{chen2022maskedimagemodelingadvances, walsh}.
The MAE was trained using the PyTorch Lightning library to enable scalable learning.
All experiments utilise the AdamW optimiser. 
For experiments using the ViT architecture, a cosine learning rate scheduler with a 10\% linear warmup is applied, whereas convolutional neural network (CNN) architectures are trained without any warmup.
We initiate two baselines for the ILD diagnosis task.
The first baseline involves creating montages from chest CT scans and applying supervised learning \cite{walsh}.
The creation of montages results in over 50,000 images.
We apply an Inception-ResNet v3 and ViT-B for this method.
We initiate the Inception-ResNet v3 with RadImageNet weights \cite{36} and the ViT-B with ChexZero weights \cite{tiu2022expert}.
The second baseline involves a modified pipeline based on Fontanellaz et al. \cite{29}, where lung patterns are segmented by training an nnU-Net \cite{Isensee2020nnUNet} in place of the original multilayer perceptron mixer network. Radiomic features, along with lung pattern percentages, are then extracted and fed into a random forest for classification.
The trained nnU-Net model is also made publicly available.
Both baselines are reported alongside the comparison against our pretrained MAE.
All experiments were performed on a single NVIDIA H100 GPU with 96 GB of VRAM. 

\begin{table}[t]
\caption{\textbf{Training configurations for all methods.}
The \textcolor{darkpastelgreen}{\textbf{green rows}} indicate our proposed method, while the \textcolor{gray}{\textbf{gray rows}} show our comparison against baselines.
Modality refers to pre-training (PT), fine-tuning (FT), and linear probing (LP). The weights of the {{Inception-ResNet and ViT-B}} models were respectively initialized with weights from RadImageNet \cite{36} and ChexZero \cite{tiu2022expert}.}
\label{tab1}
\begin{adjustbox}{max width=\linewidth, center}
\begin{tabular}{l|c|c|c|c|c|c|c|c}
\toprule
\textbf{Model} & \textbf{Modality} & \textbf{Batch Size} & \textbf{Learning Rate} & \textbf{Weight Decay} & \textbf{Iters.} & \textbf{Warm-up} & \textbf{Augmentation}\\
\midrule
\cellcolor{darkpastelgreen!25}MAE (ViT-B) & PT & 64 & 3e-4 & 5e-2 & 198,000 & \textcolor{darkpastelgreen}{\ding{51}} & \textcolor{darkpastelgreen}{\ding{51}}\\
\cellcolor{darkpastelgreen!25}MAE (ViT-B) & FT & 12 & 1e-4 & 1e-4 & 95   & \textcolor{darkpastelgreen}{\ding{51}} & \textcolor{red}{\ding{55}}\\
\cellcolor{darkpastelgreen!25}MAE  (ViT-B) & LP & 12 & 1e-2 & 1e-2 & 620  & \textcolor{darkpastelgreen}{\ding{51}} & \textcolor{red}{\ding{55}}\\
\cellcolor{gray!25}Inception-ResNet \cite{walsh} & FT & 32 & 1e-3 & 1e-4 & 11,500   & \textcolor{red}{\ding{55}} & \textcolor{red}{\ding{55}}\\
\cellcolor{gray!25}ViT-B \cite{walsh} & FT & 32 & 1e-4 & 1e-4 & 11,500  & \textcolor{darkpastelgreen}{\ding{51}} & \textcolor{red}{\ding{55}}\\
\bottomrule
\end{tabular}
\end{adjustbox}
\label{tab2}
\end{table}

\subsubsection{Evaluation metrics} 
We evaluate balanced accuracy, the average F1-score (weighted), and loss for classification.
To ensure more reliable estimates of absolute performance, we repeat data splits (70:30) five times and train five models, reporting the mean validation metric and standard deviation.


\section{Results}


In this section, we discuss experimental results and our key findings. 

\begin{table}[t]
\caption{\textbf{Binary classification results.} The \textcolor{darkpastelgreen}{\textbf{green rows}} indicate our method, while the \textcolor{gray}{\textbf{gray rows}} show our show our comparison against baselines. FT refers to full fine-tuning and LP to linear probing. 
}
\label{tab:2way}
\begin{adjustbox}{max width=\linewidth, center}
\begin{tabular}{l|c|c|c}
\toprule
\textbf{Method} & \textbf{Balanced Accuracy (\%)} & \textbf{F1 Score} & \textbf{Val Loss} \\
\midrule
\cellcolor{darkpastelgreen!25}MAE (FT) & 69.1 ± 6.4 & 0.69 ± 0.05 & 0.89 ± 0.29 \\
\cellcolor{darkpastelgreen!25}MAE (LP) & \textbf{72.3 ± 2.2}& \textbf{0.71 ± 0.02} & 0.71 ± 0.06 \\
\cellcolor{gray!25}Inception-ResNet \cite{walsh} & \textbf{72.3 ± 1.0}& \textbf{0.71 ± 0.02} & 2.62 ± 1.07 \\
\cellcolor{gray!25}ViT-B \cite{walsh} & 61.7 ± 4.8 & 0.61 ± 0.06 & 2.44 ± 0.79 \\
\cellcolor{gray!25}Radiomic Features \cite{29} & 71.7 ± 6.9 & 0.70 ± 0.08 & \textbf{0.57 ± 0.07} \\
\bottomrule
\end{tabular}
\end{adjustbox}
\end{table}

\begin{table}[t]
\caption{\textbf{Multiclass classification results.} The experimental setup and metrics are consistent with those presented in Table \ref{tab:2way}.}
\label{tab:4way}
\begin{adjustbox}{max width=\linewidth, center}
\begin{tabular}{l|c|c|c}
\toprule
\textbf{Method} & \textbf{Balanced Accuracy (\%)} & \textbf{F1 Score} & \textbf{Val Loss} \\
\midrule
\cellcolor{darkpastelgreen!25}MAE (FT)     & 43.1 ± 10.2 & 0.48 ± 0.07 & 1.78 ± 0.48 \\
\cellcolor{darkpastelgreen!25}MAE (LP)         & \textbf{49.1 ± 7.0} & \textbf{0.51 ± 0.05} & 1.45 ± 0.41 \\
\cellcolor{gray!25}Inception-ResNet \cite{walsh}     & 40.8 ± 4.7 & 0.47 ± 0.05 & 5.63 ± 2.78 \\
\cellcolor{gray!25}ViT-B \cite{walsh}                  & 28.6 ± 1.6 & 0.33 ± 0.05 & 5.01 ± 1.02 \\
\cellcolor{gray!25}Radiomic Features  \cite{29}       & 42.5 ± 7.9 & 0.48 ± 0.09 & \textbf{1.23 ± 0.12} \\
\bottomrule
\end{tabular}
\end{adjustbox}
\end{table}

\subsection{Findings}
\subsubsection{Classification Tasks}
Even with the use of AI, diagnosing ILDs from chest CT images remains challenging, due to substantial inter-reader variability, limited availability of high-quality annotated data, and the complex, heterogeneous nature of ILD patterns. 
In each result, we also add the validation loss value to measure the confidence of our model's results compared to the baseline.
Table \ref{tab:2way} compares the performance of our approaches in the binary classification task.
We observe that the MAE linear probe and baseline InceptionResNet achieved the top results, with the baseline having a smaller standard deviation.
The radiomic features are second with the lowest validation loss, but the variability is the highest of all models.
The MAE, fully fine-tuned, achieves a slightly worse result.
In Table \ref{tab:4way}, we see the results for the multiclass classification. 
The results overall are generally lower, most likely due to the overlapping of imaging features over classes.
In this case, MAE with linear probing achieves the best result at 49.1\%, outperforming the fully fine-tuned MAE, which scored 43.1\%, by an absolute margin of 6.0\%.
We found that fine-tuning the MAE led to quicker overfitting compared to the linear probing method, which in turn motivated us to train the linear probing approach for a longer duration.
ViT baseline performs worse in all cases, which is surprising since the pre-trained weights should help the learning.
Lastly, we note that the validation loss is consistently lower with MAE training compared to Walsh et al. \cite{walsh}, though the lowest overall validation loss is achieved by Fontanellaz et al. \cite{29}, albeit with consistently higher variability than the linear probing approach.

\begin{figure}[h]
\centering
\includegraphics[width=\linewidth, height=6cm, keepaspectratio]{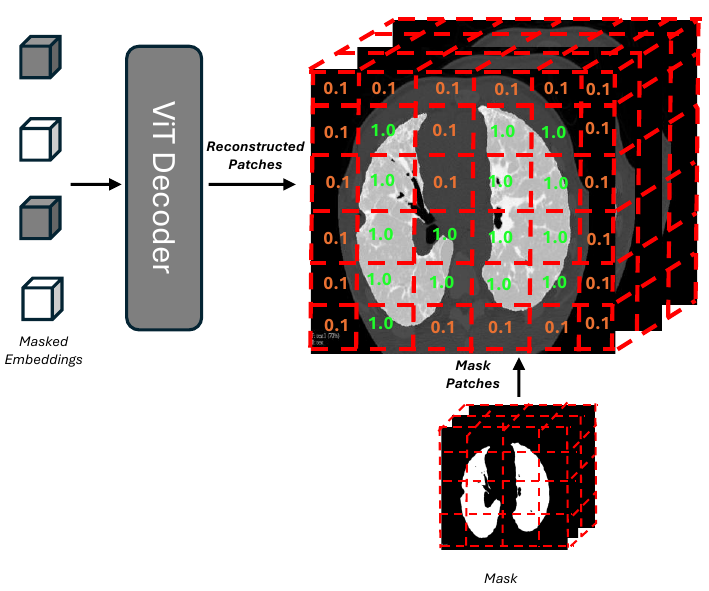}
\caption{\textbf{Lung-cavity-aware Reconstruction Loss.} The masked embeddings predicted by the ViT encoder are fed into the ViT decoder and reconstructed via the ablated loss function.
The green text denotes patches that contain at least 25\% lung voxels, the orange text shows patches that are penalised by multiplying the loss value by a lower weight (0.1 in this figure). 
}
\label{fig:loss}
\end{figure}

\subsection{Ablation study}

We perform two ablations to further analyse alternative training set-ups for MAEs on thoracic CT images.
We investigate a different patch size and a new reconstruction loss.

\subsubsection{Patch Size}
In our first ablation study, we trained the MAE using the same learning rate and weight decay but reduced the patch size to eight, resulting in 4096 patches for a 128$\times$128$\times$128 volume. 
Due to computational constraints, we set the batch size to eight and trained for 2000 epochs. 
When fine-tuning the MAE, we achieved an F1 score of 0.71 ± 0.06 and balanced accuracy of 71.0 ± 7.3 on the binary task. For the multiclass task, we obtained an F1 score of 0.45 ± 0.02 and balanced accuracy of 38.0 ± 5.6.
While the overall performance was lower, training was considerably more stable.

\subsubsection{Lung-cavity-aware Reconstruction Loss} We focus our next ablation on creating a different loss function to help reconstruct fine details within the CT scan.
Intricate lung details play a major role in diagnosing ILDs \cite{02, 05}, so it is vital that these details are reconstructed in higher resolution.
The focus on this arises because background regions in the volume can cause the mean absolute error to decrease significantly, giving the false impression that reconstruction is performing well, as shown in \cite{xiao2023delving}.
However, within the lung itself, fine details may be lost, as reconstructed voxels tend to be assigned average values, resulting in a loss of important structural information.
We show in Figure \ref{fig:loss} the overview of our loss function.
After obtaining the reconstructed patches, we patchify the scans with respect to the respective lung mask. 
For both sets of patches, we calculate the overlap with the lung mask.
If a reconstructed patch contains at least 25\% of the lung mask, we retain its index, resulting in two new sets: lung patches and non-lung patches.
We then compute the mean absolute error separately for the lung and non-lung regions. 
To reduce the influence of the non-lung area, we apply a penalty by multiplying its reconstruction error by a hyperparameter defined at the start of training, such that the loss $\mathcal{L}$ is now defined as:
$$
\mathcal{L} = \mathcal{E}_{\text{lung}} + \alpha \cdot \mathcal{E}_{\text{non-lung}}
$$
where $\mathcal{E}$ denotes the mean absolute error, and $\alpha$ is a predefined weighting hyperparameter reducing the impact of non-lung patches.
With $\alpha = 0.01$, the multiclass task achieved a balanced accuracy of 41.8 ± 4.2 and an F1 score of 0.47 ± 0.06. 
For the binary task, we obtained a balanced accuracy of 73.2 ± 4.9 and an F1 score of 0.73 ± 0.05.

\section{Discussion and Conclusion}

This study tackles the challenge of diagnosing diffuse lung diseases by leveraging self-supervised pre-training on a large, diverse chest CT dataset. 
We collected over 5,000 scans from a variety of sources, encompassing different locations, 
imaging machines, and disease types.
The diverse pathological patterns in this large dataset provided the opportunity to learn rich embeddings for ILD classification.
Training the same 3D ViT architecture in a supervised fashion yields a balanced accuracy of 50.0 ± 1.0 and an F1 score of 0.40 ± 0.08 on the binary task, and a balanced accuracy of 25.0 ± 2.0 with an F1 score of 0.24 ± 0.02 on the multiclass task.
Although ViTs are known to be difficult to train on small datasets \cite{gani2022trainvisiontransformersmallscale}, we empirically demonstrate that pretraining substantially improves performance. Specifically, our MAE-based approach outperforms traditional baselines in data-scarce settings, matching the method from \cite{walsh} on the binary task and significantly outperforming it on the multiclass task. Compared to another baseline \cite{29}, our model achieves higher performance on both tasks.
The multiclass task is challenging due to overlapping features within classes, and we note a 9.1\% increase when compared to the baseline.
The ViT Baseline most likely needs more hyperparameter tuning, but it is well known that ViT suffers from unstable training \cite{steiner2022trainvitdataaugmentation}.
The higher standard deviation in MAE results is likely due to the smaller dataset size, with only 74 scans used for training compared to over 50,000 images in the baseline.
Ablation studies showed that smaller patch sizes did not improve accuracy but led to more stable fine-tuning.
This stability, however, came at a considerable computational cost — even with 1,000 fewer training epochs, training time increased by 2.5$\times$.
Additionally, the modified loss function showed slight improvements in the binary classification task, but continued to underperform in the multiclass setting when compared to the default loss.
Future work should explore combining these approaches to further boost performance.
We also aim to explore pre-training on the lung parenchyma alone, aggregating patch representations for fine-tuning instead of relying on $\texttt{[CLS]}$ tokens, and extending this approach to tasks such as predicting forced vital capacity decline.
Further investigation is needed to understand how the additional pretraining data contributed to the MAE’s learning and its impact on downstream performance. 
We aim to understand how to select images that more effectively capture the visual features most relevant to ILD diagnosis.
A key limitation of this work is that our evaluation is conducted on a small dataset, reflecting the rarity of ILDs. 
Small datasets can cause drastic shifts in evaluation metrics.
We plan to collect and evaluate the model on an independent ILD cohort to better assess its generalisability.

\section{Acknoledgements}
This work was supported in part by the
Emergency Department and the Department of Diagnostic,
Interventional and Pediatric Radiology of Inselspital
Bern and in part by Campus Stiftung Lindenhof Bern
(SLB).
Calculations were performed on UBELIX (https://www.id.unibe.ch/hpc), the HPC cluster at the University of Bern.


%
%
%
%
\bibliographystyle{splncs04}
\bibliography{refs}

\begin{thebibliography}{10}
\providecommand{\url}[1]{\texttt{#1}}
\providecommand{\urlprefix}{URL }
\providecommand{\doi}[1]{https://doi.org/#1}

\bibitem{An2020}
An, P., Xu, S., Harmon, S.A., et~al.: Ct images in covid-19 (2020)

\bibitem{balestriero2024learningreconstructionproducesuninformative}
Balestriero, R., LeCun, Y.: Learning by reconstruction produces uninformative features for perception (2024)

\bibitem{monai}
Cardoso, M.J., Li, et~al.: Monai: An open-source framework for deep learning in healthcare (2022)

\bibitem{NEURIPS2022_db174d37}
Chen, C., Zhang, J., Xu, Y., et~al.: Why do we need large batchsizes in contrastive learning? a gradient-bias perspective (2022)

\bibitem{chen2020simpleframeworkcontrastivelearning}
Chen, T., Kornblith, S., Norouzi, M., Hinton, G.: A simple framework for contrastive learning of visual representations (2020)

\bibitem{chen2022maskedimagemodelingadvances}
Chen, Z., Agarwal, D., Aggarwal, K., et~al.: Masked image modeling advances 3d medical image analysis (2022)

\bibitem{28}
Christe, A., Peters, A., Drakopoulos, e.a.: Computer-aided diagnosis of pulmonary fibrosis using deep learning and ct images. Invest Radiol  (2019)

\bibitem{05}
Dack, E., Christe, A., Fontanellaz, M., et~al.: Artificial intelligence and interstitial lung disease: Diagnosis and prognosis. Invest Radiol  (2023)

\bibitem{dack2023empirical}
Dack, E., Brigato, L., McMurray, M., et~al.: An empirical analysis for zero-shot multi-label classification on covid-19 ct scans and uncurated reports. In: Proceedings of the IEEE/CVF International Conference on Computer Vision (2023)

\bibitem{01}
Das, S., Ayus, I., Gupta, D.: A comprehensive review of {COVID-19} detection with machine learning and deep learning techniques. Health Technol.  (2023)

\bibitem{Depeursinge2012}
Depeursinge, A., Vargas, A., Platon, e.a.: Building a reference multimedia database for interstitial lung diseases. Computerized Medical Imaging and Graphics  (2012)

\bibitem{Desai2020}
Desai, S., Baghal, A., Wongsurawat, T., et~al.: Data from chest imaging with clinical and genomic correlates representing a rural covid-19 positive population (2020)

\bibitem{Durhan2023}
Durhan, G.e.a.: Two in one: Overlapping ct findings of covid-19 and underlying lung diseases. Clinical Imaging  (2023)

\bibitem{29}
Fontanellaz, M., Christe, A., Christodoulidis, e.a.: Computer-aided diagnosis system for lung fibrosis: from the effect of radiomic features and multi-layer-perceptron mixers to pre-clinical evaluation. IEEE Access  (2024)

\bibitem{gani2022trainvisiontransformersmallscale}
Gani, H., Naseer, M., Yaqub, M.: How to train vision transformer on small-scale datasets? (2022)

\bibitem{he2021maskedautoencodersscalablevision}
He, K., Chen, X., Xie, S., et~al.: Masked autoencoders are scalable vision learners (2021)

\bibitem{lungmask}
Hofmanninger, J., Prayer, F., Pan, J., et~al.: Automatic lung segmentation in routine imaging is primarily a data diversity problem, not a methodology problem. Eur Radiol Exp  (2020)

\bibitem{Huang2023}
Huang, S., Pareek, A., Jensen, M., et~al.: Self-supervised learning for medical image classification: a systematic review and implementation guidelines. npj Digital Medicine  (2023)

\bibitem{Isensee2020nnUNet}
Isensee, F., Jaeger, P.F., Kohl, e.a.: {nnU-Net}: a self-configuring method for deep learning-based biomedical image segmentation. Nature Methods  (2020)

\bibitem{Kiser2020}
Kiser, K.J.e.a.: Data from the thoracic volume and pleural effusion segmentations in diseased lungs for benchmarking chest ct processing pipelines (plethora) (2020)

\bibitem{Kun_Training_MICCAI2024}
Kunanbayev, K., Shen, V., Kim, D.S.: { Training ViT with Limited Data for Alzheimer’s Disease Classification: an Empirical Study }. In: MICCAI (2024)

\bibitem{Lynch2018}
Lynch, D.A., Sverzellati, N., Travis, W.D., et~al.: Diagnostic criteria for idiopathic pulmonary fibrosis: A fleischner society white paper. Lancet Respiratory Medicine  (2018)

\bibitem{MP-COVID-19-SegBenchmark}
Ma, J., Wang, Y., An, X., et~al.: Towards data-efficient learning: A benchmark for covid-19 ct lung and infection segmentation. Medical Physics  (2021)

\bibitem{36}
Mei, X., Liu, Z., Robson, P.M., et~al.: Radimagenet: An open radiologic deep learning research dataset for effective transfer learning. Radiology: Artificial Intelligence  (2022)

\bibitem{Morozov2020}
Morozov, S.P., Andreychenko, A.E., Pavlov, N.A., et~al.: {MosMedData}: Chest {CT} scans with {COVID-19} related findings dataset (2020)

\bibitem{04}
Ng, Y.A., et~al.: A novel workflow for the safe and effective integration of ai as supporting reader in double reading breast cancer screening: A large-scale retrospective evaluation. medRxiv  (2022)

\bibitem{Oikonomou2021}
Oikonomou, A., et~al., K.P.: {COVID-CT-MD: COVID-19 Computed Tomography Scan Dataset Applicable in Machine Learning and Deep Learning}  (2021)

\bibitem{12}
Pratella, D., Ait-El-Mkadem~Saadi, S., Bannwarth, S., et~al.: A survey of autoencoder algorithms to pave the diagnosis of rare diseases. International Journal of Molecular Sciences  (2021)

\bibitem{055}
Rea, G., Sverzellati, N., Bocchino, M., et~al.: Beyond visual interpretation: Quantitative analysis and artificial intelligence in interstitial lung disease diagnosis "expanding horizons in radiology". Diagnostics (Basel)  (2023)

\bibitem{stoic}
Revel, Marie-Pierre, e.a.: Study of thoracic ct in covid-19: The stoic project. Radiology  (2021)

\bibitem{SAHA2024779}
Saha, S., Kumar, A., Nandi, D.: Vit-ild: A vision transformer-based neural network for detection of interstitial lung disease from ct images. Procedia Computer Science  (2024)

\bibitem{34}
Shahin, A., Wegworth, C., David, Estes, E., et~al.: Osic pulmonary fibrosis progression. Kaggle  (2020)

\bibitem{steiner2022trainvitdataaugmentation}
Steiner, A., Kolesnikov, A., Zhai, X., et~al.: How to train your vit? data, augmentation, and regularization in vision transformers (2022)

\bibitem{tiu2022expert}
Tiu, E., Talius, E., Patel, P., et~al.: Expert-level detection of pathologies from unannotated chest x-ray images via self-supervised learning. Nature Biomedical Engineering  (2022)

\bibitem{02}
Trusculescu, A., Manolescu, D., Tudorache, E., et~al.: Deep learning in interstitial lung disease-how long until daily practice. Eur Radiol  (2020)

\bibitem{walsh}
Walsh, S., Calandriello, L., Silva, M., et~al.: Deep learning for classifying fibrotic lung disease on high-resolution computed tomography: a case-cohort study. Lancet Respir Med  (2018)

\bibitem{sabir2023fibrovit}
Waseem~Sabir, M., Farhan, M., Almalki, N.S., et~al.: {FibroVit—Vision transformer-based framework for detection and classification of pulmonary fibrosis from chest CT images}. Frontiers in Medicine  (2023)

\bibitem{xiao2023delving}
Xiao, J., Bai, Y., Yuille, e.a.: Delving into masked autoencoders for multi-label thorax disease classification. In: Proceedings of the IEEE/CVF Winter Conference on Applications of Computer Vision (2023)

\bibitem{covid-vit}
Xiaohong~Gao, Yu~Qian, A.G.: Covid-vit: Classification of covid-19 from ct chest images based on vision transformer models. arXiv:2107.01682  (2021)

\bibitem{Zaffino2021}
Zaffino, P., Marzullo, A., Moccia, S., et~al.: An open-source covid-19 ct dataset with automatic lung tissue classification for radiomics. Bioengineering  (2021)

\bibitem{zhou2023selfpretrainingmaskedautoencoders}
Zhou, L., Liu, H., Bae, J., et~al.: Self pre-training with masked autoencoders for medical image classification and segmentation (2023)

\end{thebibliography}
\end{document}